# Understanding the intrinsic framework of the Hall-Petch relationship of metals from the view of the electronic-structure level

Xin Li, Wang Gao∗, Qing Jiang

Key Laboratory of Automobile Materials, Ministry of Education, Department of Materials Science and Engineering, Jilin University, 130022, Changchun, China

∗wgao@jlu.edu.cn

**Abstract:** The relationship between grain size and yield strength of metals follows the Hall-Petch relationship $\sigma = \sigma_0 + kd^{-0.5}$; however, the specific physical factors that affect the coefficients $\sigma_0$ and $k$ of this relationship remain unclear. Here we propose the intrinsic descriptors to determine the Hall-Petch relation across different metals and alloys. Inspired by the tight-binding theory, we find that $\sigma_0$ strongly depends on the group and period number, the valence-electron number and electronegativity, while $k$ is determined by the cohesive energy. Our framework establishes a predictive structure-property relationship for the size-dependent yield strength of various metals, and unravels that both the coefficients of the Hall-Petch relationship physically originate from the d-band properties. This novel correlation provides a new perspective for understanding the mechanical strength of metals, which is useful for the design of high-performance materials.

# 1. Introduction

The Hall-Petch relation has been widely used in determining the yield strength of polycrystalline metals with the change of grain sizes, which is expressed as $\sigma = \sigma_0 + kd^{-0.5}$ [1, 2]. Here $\sigma$ represents the yield strength, $d$ corresponds to the grain size, and $\sigma_0$ and $k$ are the material-dependent parameters referring to the lattice frictional stress and the Hall-Petch coefficient respectively [3]. Despite the practical success of the Hall-Petch relation in engineering, unraveling the underlying physical origin of $\sigma_0$ and $k$ and subsequently determining the yield strength have remained formidable challenges in the field of metallic materials [4, 5].

The parameters $\sigma_0$ and $k$ are typically fitted by the experimentally measured yield stress in the previous studies. However, the values of $\sigma_0$ and $k$ can vary significantly depending on the specific experimental techniques and testing methods employed. For example, hardness measurements usually overestimate the flow stress compared with other mechanical testing approaches [4, 5], and the X-ray diffraction (XRD) patterns tend to underestimate the grain dimensions compared with transmission electron microscopy (TEM) observations in general [5]. On the other hand, the thermomechanical processing techniques used for sample preparation, such as swaging, rolling, or forging, followed by recrystallization annealing, can introduce deviation in yield strength and secondary effects on grain-size strengthening behavior [4, 6]. All these different methods lead to deviations in the fitting parameters of the Hall-Petch relationship. Besides the experimental measurements, many efforts have been made through theoretical research to understand the Hall-Petch relation like the dislocation-pile-up models [1, 7-11] geometrically-necessary-dislocations models [3, 12], slip-distance models [13, 14], and other models inspired by the ideas of work hardening [15, 16]. All these theoretical studies promote the development of the Hall-Petch relation and deepen the understanding of its fundamental mechanism. However, these theories focus on the motion of grain boundaries and dislocations from an atomic-scale level, leading to the models containing undetermined material-dependent parameters and lacking the electronic-structure characteristics of metals. The developed models considering the coarse-grain length scale [17] and based on the machine learning approach propose the novel predictive Hall-Petch relation [18] for diverse metals, but need the macroscopic properties and the factors calculated by first-principles calculations and molecular dynamics simulations. Therefore, it remains the challenge to provide a universal scheme of the Hall-Petch relation with the intrinsic characteristics, which uncovers the physical picture of the macroscopic properties directly from the electronic-structure level.

We propose a novel intrinsic framework for the Hall-Petch relation: the lattice-frictional term $\sigma_0$ can be determined by the descriptor based on the group and period number, valence-electron number, and electronegativity, while the Hall-Petch coefficient $k$ strongly depends on the cohesive energy of metals. Unlike previous theories that primarily focus on the movement of grain boundaries and dislocations and correlate yield strength with the macroscopic feature of metals, our scheme elucidates the physical origin of the Hall-Petch relation from the electronic-level perspective through the tight-binding theory, providing a predictive tool for determining the yield stress of metals and a new insight into understanding the mechanism of material strengthening.

## 2. Methods

We collect the size-dependent yield strength of 20 metals commonly used in structural materials from the literature [4, 5]. The sample preparation of these metals contains swaging, rolling or forging, followed by recrystallization anneals and other generally used specimen preparation approaches. Various testing methods are also involved in the collective data, including the tension and compression tests as well as the Vickers and nanoindentation hardness measurements. The latter has been divided by a Tabor factor of 3 to convert to strength measurements [19]. Most of the flow stresses are measured at plastic strains of order 0.2%. All these experimental data have been used and verified in the previous studies about the Hall-Petch relation [4, 5]. Based on the collective data, we fit the $\sigma_0$ and $k$ for each metal respectively. Note that the hardness measurements may have offset values above the yield strength ones since they include some plastic strain and introduce the work-hardening effect [4, 20, 21]. To compare the influence of hardness and tension/compression tests in the Hall-Petch relation, we further consider the different experimental approaches separately during the data fitting. Both of the measured data satisfy the Hall-Petch relation, but $k$ and $\sigma_0$ are distinct across different test approaches. Therefore, we address the data by adopting three fitting strategies to obtain the $\sigma_0$ and $k$: the data from the individual tension and compression test, the ones from the individual hardness test and the ones from both two tests. Taking the yield strength of Ti as the example (Fig. 1), $\sigma_0$ and $k$ of Ti are 321.28 and 170.01 by fitting from the data of the hardness measurements, while 119.67 and 284.45 from the data of tension and compression tests, and 192.56 and 230.18 from all the data of both the measurements [4]. All the regression coefficients from these three fitting approaches are above 0.9, demonstrating that the coefficients of the Hall-Petch relation are strongly dependent on the measurement method.

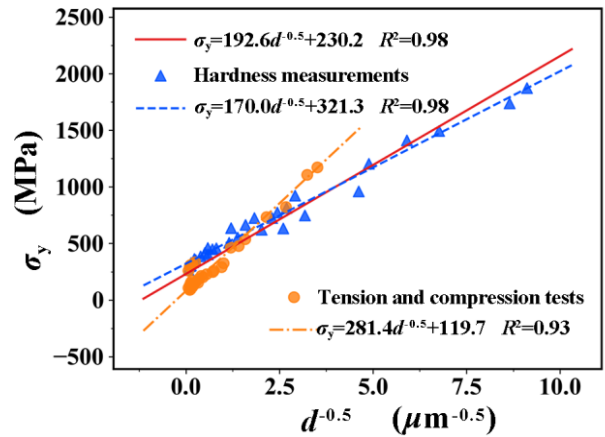

Figure 1 The yield strength as the linear function of $d^{-0.5}$ fitting from the data using hardness measurements, tension and compression tests and both two measurements [4].

## 3. Results and Discussion

### 3.1 Lattice friction term $\sigma_0$ in the Hall-Petch relationship of metals.

It is widely known that the $\sigma_0$ is commonly attributed to lattice-friction resistance, which signifies the resistance encountered by dislocations as they move within grains, while the slope $k$ relates to the grain boundary diffusion and the interactions between dislocations and grain boundary[4, 5]. We first focus on the lattice friction term $\sigma_0$ of different metals.

The previous models demonstrate that as the dislocation moves through the lattice, its energy fluctuates with period Burgers vector $b$. The atomic energy potential $U(x)$ at dislocation can be expressed as Fourier series,

$$U(x) = U_0 + U_1\cos(2\pi x/b) + U_2\cos(4\pi x/b) + U_3\cos(8\pi x/b) + ... \quad (1)$$

Here the $U_0$, $U_1$, … depend on different metals. Taking the first-order approximation of the Fourier series of energy potential, the interatomic force can be expressed as

$$\sigma(x) = N_s \frac{dU}{dx} = N_s U_1 \frac{2\pi}{b} \sin(\frac{2\pi x}{b}) \quad (2)$$

Here $N_s$ is the number of atoms per unit area on the cutting surface. This force essentially reflects the tensile stress required to move the atoms at the stable equilibrium position, in order to lead to the dislocation motion. The maximum tensile stress is shown as,

$$\sigma_{max} = N_s U_1 \frac{2\pi}{b} \qquad (3)$$

In the perfect metal crystal, the parameter $U_1$ depends on the cohesive energy $E_{coh}$. However, the coordination numbers (*CN*) of the atoms around the dislocation containing the strain effects thus vary dramatically from those near the equilibrium position. The bond breaking and forming cause the change of d-band center ($\varepsilon_d$) and d-band width ($W_d$) of the transition-metal atoms around the dislocation according to tight-binding approximation [22],

$$\sigma_{max} = N_s \frac{2\pi}{b} E_{coh} f(\Delta\varepsilon_d, \Delta W_d) \qquad (4)$$

It is noteworthy that $\sigma_{max}$ in Eq. (4) reflects the maximum tensile stress of the metals with defects, especially with the dislocations. The motion of dislocations causes the variation in the geometric environment around the atoms, corresponding to the atomic bond breaking and forming as well as the change of d-band center and width. As bond breaking and forming are directly related to the valence electron number ($S_v$) and electronegativity ($\chi$), we adopt the descriptor $\psi = \frac{S_v^2}{\chi}$ that reflects the d-band center and describes the surface adsorption well to capture the bond breaking and forming. Moreover, $W_d$ depends on the group and period number ($N_g$ and $N_p$) as $W_d$ scales with the $N_g$ for the same $N_p$ (see Fig. 2). Therefore, we introduce the descriptor $Ж = \left(\frac{N_p}{4}\right)^{(\sqrt{N_g}-3)} \times \frac{S_v^2}{\chi}$, which contains the coupling effects of $\varepsilon_d$ and $W_d$ [23, 24], to capture the binding properties of atoms around dislocations and study the lattice friction term $\sigma_0$.

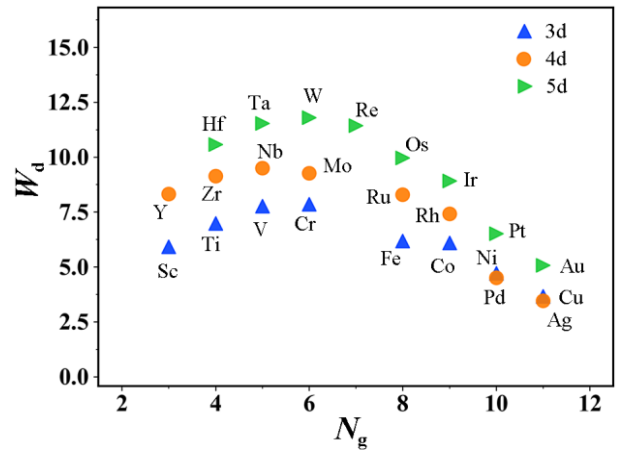

Figure 2 The correlation between d-band width $W_d$ and the group and period number $N_g$ and $N_p$ of transition metals.

Remarkably, the descriptor $Ж$ describes the trend of $\sigma_0$ in a broken-line relationship for different metals including the transition metals and some main-group metals (Fig. 3). This relationship is solid for the data from various fitting approaches, such as the data from the individual tension and compression test (Fig. 3b), from the individual hardness test (Fig. 3c) as well as from both two tests (Fig. 3a). Moreover, we also use the $\sigma_0$ values that have been fitted in the literature and simply average them [4, 5]. These average $\sigma_0$ also exhibit similar broken-line relations with Ж for different metals (Fig. 3d). All these results demonstrate that the material-dependent lattice friction terms of the Hall-Petch relationship are well determined by Ж, which reflects the d-band properties of the bond breaking and forming, regardless of the experimental tests and fitting approaches.

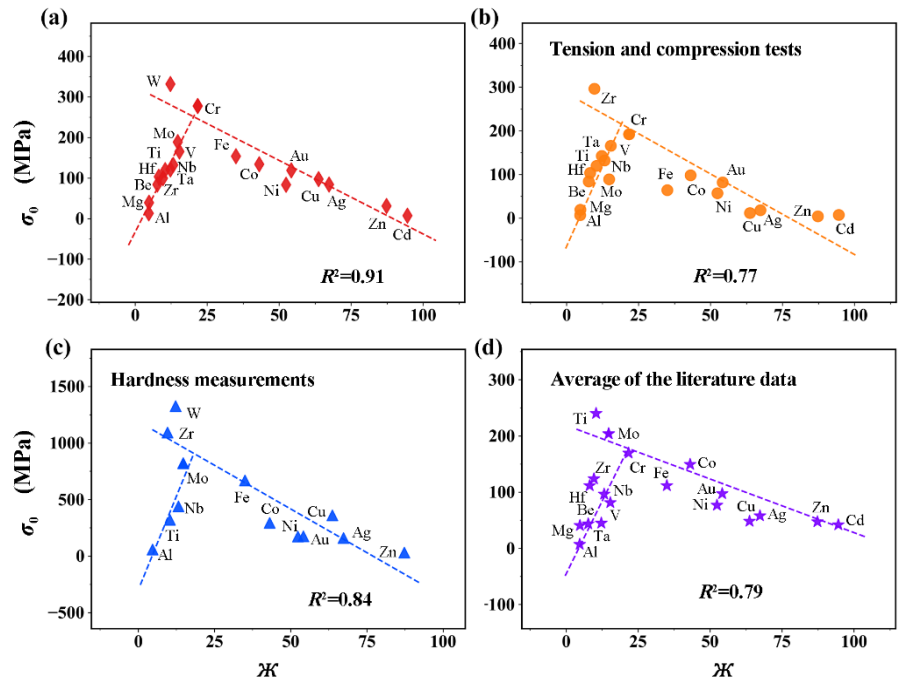

Figure 3 The lattice friction term $\sigma_0$ as the function of the descriptor $Ж$ for different metals [4]. (a) $\sigma_0$ fitting from the available data using hardness measurements and tension and compression tests. (b) $\sigma_0$ fitting from the available data only using tension and compression tests. (c) $\sigma_0$ fitting from the available data only using hardness measurements. (d) The average $\sigma_0$ of the literature data [4,5].

### 3.2 Hall-Petch coefficient *k* in the Hall-Petch relationship of metals.

We turn to study the Hall-Petch coefficient *k* of the Hall-Petch relation, which depends on the ability of dislocation movement passing through the grain boundary. The dislocations glide at the grain boundary, causing the grain boundary sliding and accumulating the stress at triple junctions. The grain boundary diffusion thus acts as an important factor in measuring the Hall-Petch coefficient based on the grain-boundary-sliding models [25]. The activation energy of grain boundary diffusion for different metals can be estimated by the melting temperature of materials [26]. These results imply that the grain-boundary-sliding properties can be reflected by the cohesive properties of metals, since the cohesive forces holding atoms in a crystal lattice must be overcome in the melting process [27]. Therefore, we try to correlate the cohesive energy ($E_{coh}$) with the Hall-Petch coefficients *k*. We find that the Hall-Petch coefficients *k* of different metals exhibit a well-linear relationship with the square of the cohesive energy ($E_{coh}^2$), applicable to the different data obtained from distinct measurements (Fig. 4a-c) or the average data of literature (Fig. 4d). The square of the cohesive energies can also linearly determine the activation energies of grain boundary diffusion and the melting temperature of different metals (Fig. 5a), further demonstrating that the cohesive properties determine the ability of grain boundary diffusion, and thus impact the Hall-Petch coefficients of different metals.

## 3.3 The origin of the coefficients in the Hall-Petch relation for different metals.

Overall, we discuss the coefficients of the Hall-Petch relation for different metals from the view of bond breaking and formation, which is robust for various experimental tests. The lattice friction term $\sigma_0$ reflects the thermodynamic characteristic of dislocation stability, which can be determined by the electronic descriptor $\mathcal{K}$ based on the group and period number, valence electron number, and electronegativity. The grain boundary barrier term $k$ reflects the dynamic properties of the grain boundary diffusion, which can be determined by the square of cohesive energy. Inspired by the tight-binding approximation, both of the two descriptors reflect the d-band properties—Ж as a linear function of the d-band upper edge ($\varepsilon_d+W_d/2$) and cohesive energy as a linear function of the d-band width (Fig. 5b and 5c) [23, 24]. The d-band center ($\varepsilon_d$) is defined as the first moment of the projected d-band density of states as [28]

$$\varepsilon_d = \frac{\int \rho E dE}{\int \rho dE} \tag{5}$$

where $\rho$ represents the density of states and $E$ is the energy of the states. The d-band width ($W_d$) is defined as the second moment of the projected d-band density of states as [28]

$$W_d^2 = \frac{\int \rho(E-\varepsilon_d)^2 dE}{\int \rho dE} \tag{6}$$

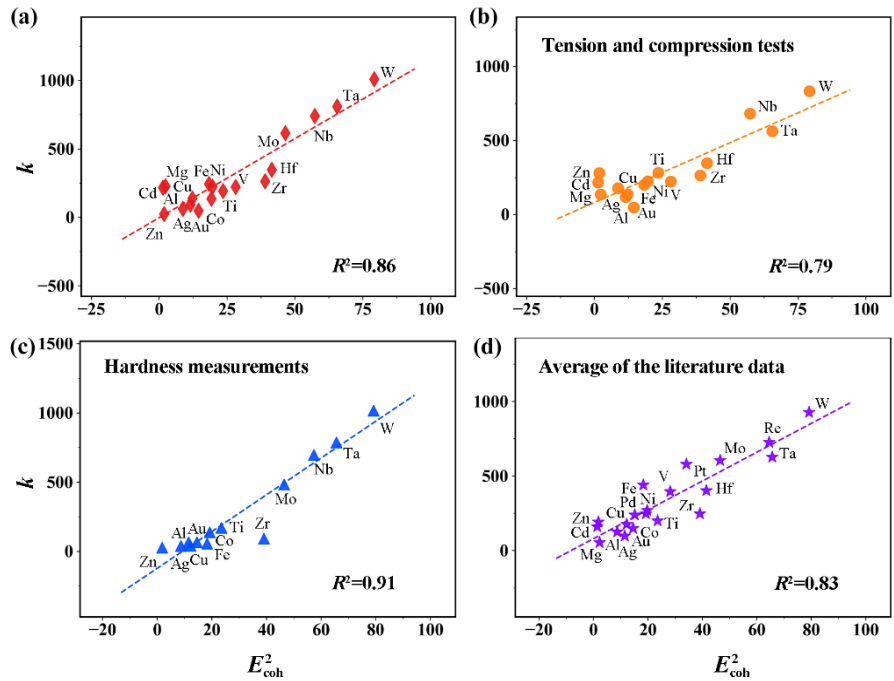

Figure 4 The Hall-Petch coefficient $k$ as the function of the square of cohesive energy $E_{coh}$ for different metals. (a) $k$ fitting from the available data using hardness measurements and tension and compression tests. (b) $k$ fitting from the available data only using tension and compression tests. (c) $k$ fitting from the available data only using hardness measurements. (d) The average $k$ of the literature data [4,5].

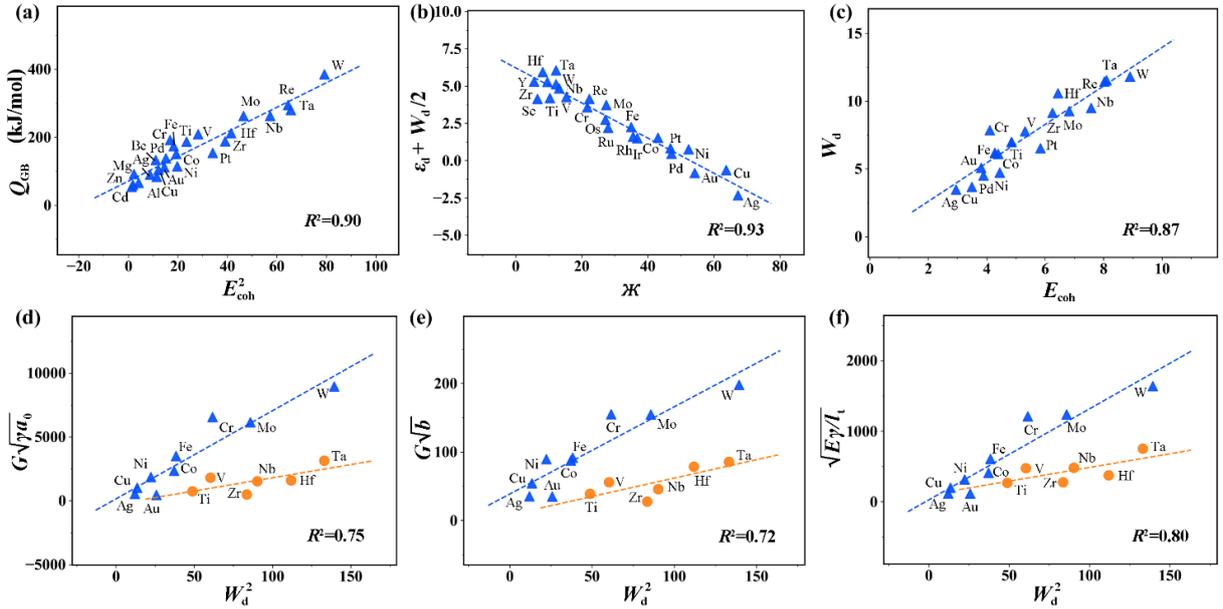

Figure 5 (a) The linear correlation between the activation energies of grain boundary diffusion and the square of $E_{coh}$ [5]. (b) Ж as a linear function of the upper edge of d-band. (c) The cohesive energy $E_{coh}$ as a linear function of the d-band width $W_d$. The linear correlation between $W_d^2$ with (d) $G\sqrt{\gamma_{GB}a_0}$, (e) $G\sqrt{b}$ and (f) $\sqrt{E\gamma_{GB}/l_t}$ [18].

which naturally explains the square term of cohesive energy ($E_{coh}^2$) in describing the Hall-Petch coefficient $k$. Combined with d-band center and width, the upper edge of d-band $\varepsilon_d + W_d/2$ can also effectively reflect the position and filling of anti-bonding states. All these findings imply that $k$ and $\sigma_0$ of the Hall-Petch relation for different metals both originate from the d-band properties: $\sigma_0$ depends on the d-band upper edge and $k$ is strongly related to the second moment of the projected d-band density of states. Moreover, we use the intrinsic properties to substitute the parameters of the d-band density of states obtained by electronic structure calculations, which can be also easily extended to the main-group metals. It is noteworthy that Cr exhibits a large Hall-Petch coefficient $k$ and acts as the outlier in the relationship between $k$ and $E_{coh}^2$. The potential reason is that the tendency of the cohesive energy and d-band width in Cr is distinct from that in other half-d-occupation metals such as Mo and W. According to the tight-binding approximation, a half d-occupation corresponds to a larger d-band width (of the metals in the same period) and thus a larger cohesive energy [27]. Cr is a special case that it exhibits a large d-band width due to the half d-occupation but a relatively small cohesive energy due to the magnetic properties, leading to a small predictive Hall-Petch coefficient $k$ based on the correlation between $k$ and $E_{coh}^2$.

The origins of Hall-Petch coefficient $k$ and the lattice friction term $\sigma_0$, from d-band width and the upper edge of d-band, are also compatible with previously proposed models. The motion of grain-boundary and dislocation are the key factors in determining the Hall-Petch coefficient $k$, which strongly depends on the shear modulus $G$, Burgers vectors $b$, lattice constant $a_0$, grain-boundary energy $\gamma_{GB}$, linear thermal expansion $l_t$ and other parameters based on the pile-up models and geometrically necessary dislocations (GND). These models derive the results as $k \propto G\sqrt{b}$ [12, 29], $k \propto G\sqrt{\gamma_{GB} a_0}$ [30], and $k \propto \sqrt{E\gamma_{GB} l_t}$ [18]. These parameters that reflect the Hall-Petch coefficient *k* are all linearly correlated with the square of d-band width. The two linear correlations between d-band width and the parameters of previous models reflect the fact that the d-band width of the transition metals follows the broken-line relationship with the group number, and reaches the turning point at the metals with half-d-band occupation (see Fig. 2). Therefore, these parameters vary dramatically between the transition metals with more than half-d-band occupation (Cr, Mo, W, Fe, Co and Ni), while change slowly between those with lower d-band occupation (Ti, Zr, Hf, V, Nb and Ta). Nevertheless, our results demonstrate that the Hall-Petch coefficient *k*, influenced by the grain-boundary diffusion, originates from the second moment of d-band density of states, which provides a novel physical picture to understand the Hall-Petch relationship.

### 3.4 The prediction of size-dependent yield strength of metals.

Based on the two intrinsic descriptors, our models can predict the size-dependent yield strength of various metals by combining with the Hall-Petch relationships. The mean absolute error (MAE) is about 69.4 MPa (1.3%), as illustrated in Fig. 6, which is comparable to the models based on machine learning methods [18], and smaller than the previous pile-up models [29] and recent theories [30, 31] with the regression coefficient of about 0.8 [18] and MAE of more than 100 MPa. Our scheme exhibits a wide application of the yield-strength estimation compared with the previous studies, which contain more than 600 data, including 13 metals that are commonly used as structural materials, with a wide range of grain sizes about 6 nm ~ 3.85 mm, and also contain distinct experimental testing approaches such as tension and compression tests and hardness measurements.

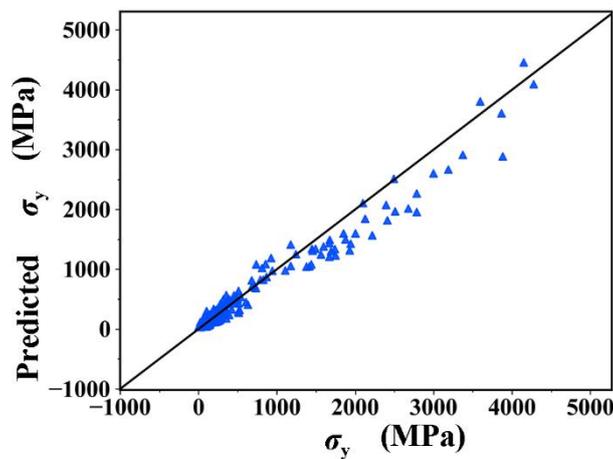

Figure 6 The comparison between the grain-size-dependent yield strength and the prediction values based on our intrinsic descriptors.

### 3.5 The generalization of our descriptor to alloys.

We attempt to expand our descriptor to multiple-elemental alloys, such as the medium- and high-entropy alloys (MEAs and HEAs). Although these alloys exhibit severe lattice distortion with uneven distribution, the solute/dislocation interaction and the motion of grain boundary are also attributed to the bond breaking and forming, and reflect the mean-field effects in general [32]. Therefore, we introduce the rule of mixture estimate (namely the average weighted by atomic concentration) into our descriptor as

$$E_{\text{coh,alloy}}^2 = \sum_{i=1}^{N} E_{\text{coh},i}^2 c_i \qquad (7)$$

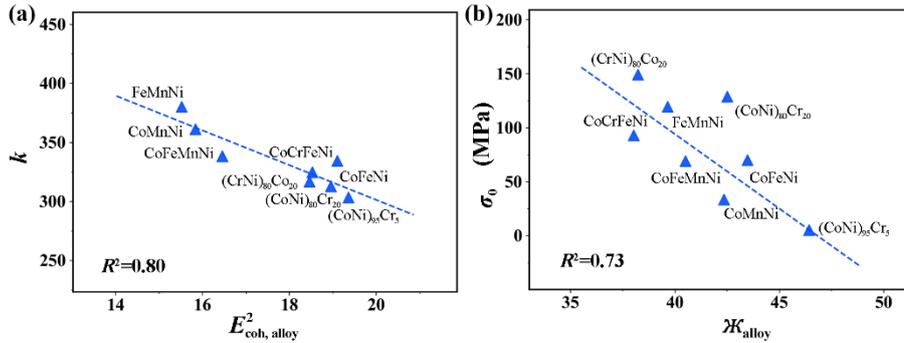

Figure 7 The linear correlation of (a) $E_{\text{coh,alloy}}^2$ vs $k$ and (b) $\mathcal{K}_{\text{alloy}}$ vs $\sigma_0$ in MEAs and HEAs [32].

$$\mathcal{K}_{\text{alloy}} = \sum_{i=1}^{N} \mathcal{K}_i c_i \qquad (8)$$

Here, $N$ is the number of elements in the multiple-component alloys, $\mathcal{K}_i$ and $E_{\text{coh},i}$ are the electronic descriptors and cohesive energies of the $i$th elements and $c_i$ is the concentration of the $i$th elements.

We find that $E_{\text{coh,alloy}}^2$ show the good linear correlation with the Hall-Petch coefficient of Fe-Co-Ni-Cr-Mn MEAs and HEAs with different elemental combinations and concentrations (Fig. 7a). The regression coefficient of $k$ vs $E_{\text{coh,alloy}}^2$ are more than 0.8, demonstrating that the Hall-Petch coefficient $k$ of MEAs and HEAs also originate from the second moment of d-band density of states combined with the mean-field effects. The lattice friction terms $\sigma_0$ of MEAs and HEAs also exhibit the linear relationship with the electronic descriptor $\mathcal{K}_{\text{alloy}}$, as illustrated in Fig. 7b. All these results demonstrate our descriptors are applicable to determine the coefficients of the Hall-Petch relationship for different metals, providing new direction and insights in discussing Hall-Petch relationship from the electronic-structure level.

## 4. Conclusions

In summary, we unravel a novel and general physical picture for the Hall-Petch relationship of metals from the electronic-level perspective, by identifying two descriptors to determine the parameters in the Hall-Petch relationship. The lattice friction term $\sigma_0$ is determined by the descriptor $\mathcal{K}$ based on the group and period number, valence electron number and electronegativity, while the grain boundary barrier term $k$ strongly correlates with the cohesive energy. The two descriptors are robust for different experimental approaches, which not only reflect the thermodynamic and dynamic characteristics of the classical models, but also uncover a novel physical picture: the coefficients of the Hall-Petch relationship can be attributed to the first and second moment of the projected d-band density of states. Moreover, the descriptor combined with the Hall-Petch relationship can accurately predict the size-dependent yield strength for transition metals and main-group metals. Our descriptors are also applicable to alloys in combination with the rule of mixture estimate, showing the generalization of our framework based on the perspective of electronic structure. All these findings provide a predictive tool for mechanical properties and novel physical guidance for the correlation between micro-bond properties and macro properties of metals.

## 5. Acknowledgments

The authors are thankful for the support from the National Natural Science Foundation of China (Nos. 22173034, 11974128, 52130101), the Opening Project of State Key Laboratory of High Performance Ceramics and Superfine Microstructure (SKL202206SIC), the Program of Innovative Research Team (in Science and Technology) in University of Jilin Province, the Program for JLU (Jilin University) Science and Technology Innovative Research Team (No. 2017TD-09), the Fundamental Research Funds for the Central Universities, and the computing resources of the High Performance Computing Center of Jilin University, China.

**References**

[1] E.O. Hall, The deformation and ageing of mild steel: III discussion of results, Proc. Phys. Soc., Sec. B 64 (1951) 747-755.

[2] N.J. Petch, The cleavage strength of polycrystals, J. Iron Steel Res. Int. 19 (1953) 25-28.

[3] V. Bata, E.V. Pereloma, An alternative physical explanation of the Hall-Petch

relation, Acta Mater. 52 (2004) 657-665.

[4] Z.C. Cordero, B.E. Knight, C.A. Schuh, Six decades of the Hall-Petch effect - a survey of grain-size strengthening studies on pure metals, Int. Mater. Rev. 61 (2016) 495-512.

[5] R.B. Figueiredo, M. Kawasaki, T.G. Langdon, Seventy years of Hall-Petch, ninety years of superplasticity and a generalized approach to the effect of grain size on flow stress, Prog. Mater Sci. 137 (2023) 101131.

[6] Y. Wang, H. Choo, Influence of texture on Hall-Petch relationships in an Mg alloy, Acta Mater. 81 (2014) 83-97.

[7] A.H. Cottrell, Theory of brittle fracture in steel and similar metals, Trans. Am. Inst. Metall. Eng 212 (1958) 192-203.

[8] R.W. Armstrong, I. Codd, R.M. Douthwaite, N.J. Petch, The plastic deformation of polycrystalline aggregates, Philos. Mag. A 7 (1962) 45-58.

[9] E. Smith, P.J. Worthington, The effect of orientation on the grain size dependence of the yield strength of metals, Philos. Mag. 9 (1964) 211-216.

[10] A. Navarro, E.R. De Los Rios, An alternative model of the blocking of dislocations at grain boundaries, Philos. Mag. A 57 (1988) 37-42.

[11] A.A. Nazarov, On the pile-up model of the grain size-yield stress relation for nanocrystals, Scripta Mater. 34 (1996) 697-701.

[12] M.F. Ashby, The deformation of plastically non-homogeneous materials, Philos. Mag. 21 (1970) 399-424.

[13] H. Conrad, Effect of grain size on the lower yield and flow stress of iron and steel, Acta Metall. 11 (1963) 75-77.

[14] J.D. Meakin, N.J. Petch, Strain-hardening of polycrystals: the alpha-brasses, Philos. Mag. 29 (1974) 1149-1156.

[15] A.W. Thompson, M.I. Baskes, W.F. Flanagan, The dependence of polycrystal work hardening on grain size, Acta Metall. 21 (1973) 1017-1028.

[16] M.A. Meyersm, E. Ashworth, A model for the effect of grain size on the yield stress of metals, Philos. Mag. A 46 (1982) 737-759.

[17] C.V. Di Leo, J.J. Rimoli, New perspectives on the grain-size dependent yield strength of polycrystalline metals, Scripta Mater. 166 (2019) 149-153.

[18] L. Jiang, H.D. Fu, H.T. Zhang, J.X. Xie, Physical mechanism interpretation of polycrystalline metals' yield strength via a data-driven method: A novel Hall-Petch relationship, Acta Mater. 231 (2022).

[19] D. Tabor, A simple theory of static and dynamic hardness, Proc. R. Soc. Lon. Ser. A-Math. Phys. Sci. 192 (1948) 247-274.

[20] M.J. Marcinkowski, H.A. Lipsitt, The plastic deformation of chromium at low temperatures, Acta Metall. 10 (1962) 95-111.

[21] C.P. Brittain, R.W. Armstrong, G.C. Smith, Hall-petch dependence for ultrafine grain size electrodeposited chromium, Scripta Metall. 19 (1985) 89-91.

[22] W.A. Harrison, Electronic structure and the properties of solids, Dover Publications, Inc., New York, 1989.

[23] W. Gao, Y. Chen, B. Li, S.P. Liu, X. Liu, Q. Jiang, Determining the adsorption energies of small molecules with the intrinsic properties of adsorbates and substrates, Nat. Commun. 11 (2020) 1196.

[24] B. Li, X. Li, W. Gao, Q. Jiang, An effective scheme to determine surface energy and its relation with adsorption energy, Acta Mater. 212 (2021) 116895.

[25] R.B. Figueiredo, T.G. Langdon, Deformation mechanisms in ultrafine-grained metals with an emphasis on the Hall-Petch relationship and strain rate sensitivity, J. Mater. Res. Tech. 14 (2021) 137-159.

[26] W. Gust, S. Mayer, A. Bögel, B. Predel, Generalized representation of grain boundary self-diffusion data, J. Phys. Colloques 46 (1985) 537-544.

[27] M.A. Turchanin, P.G. Agraval, Cohesive energy, properties, and formation energy of transition metal alloys, Powder Metall. Met. Ceram. 47 (2008) 26-39.

[28] R. Drautz, D.G. Pettifor, Valence-dependent analytic bond-order potential for transition metals, Phys. Rev. B 74 (2006) 174117.

[29] J.C.M. Li, Y.T. Chou, The role of dislocations in the flow stress grain size relationships, Metall. Mater. Trans. B 1 (1970).

[30] J. Luo, Z.R. Wang, On the physical meaning of the Hall-Petch constant, Adv. Mater. Res. 15 (2007) 643-648.

[31] D.J. Dunstan, B.A. J., Grain size dependence of the strength of metals: The Hall–Petch effect does not scale as the inverse square root of grain size, Int. J. Plast. 53 (2014) 56-65.

[32] S. Yoshida, T. Ikeuchi, T. Bhattacharjee, Y. Bai, A. Shibata, N. Tsuji, Effect of elemental combination on friction stress and Hall-Petch relationship in face-centered cubic high / medium entropy alloys, Acta Mater. 171 (2019) 201-215.